\documentclass[a4paper]{jpconf}
\usepackage{graphicx}
\usepackage{iopams}

\begin{document}
\title{Incoherent bremsstrahlung in flat and bent crystal}

\author{N F Shul'ga$^1$, V V Syshchenko$^2$ and A I Tarnovsky$^2$}

\address{$^1$ Akhiezer Institute for Theoretical Physics of the NSC
``KIPT'', Akademicheskaya Street, 1, Kharkov 61108, Ukraine}

\address{$^2$ Belgorod State University, Pobedy Street, 85, Belgorod 308015, Russian Federation}

\ead{shulga@kipt.kharkov.ua, syshch@bsu.edu.ru, syshch@yandex.ru}

\begin{abstract}
The bremsstrahlung cross section for relativistic electrons in a
crystal is split into the sum of coherent and incoherent parts
(the last is due to a thermal motion of atoms in the crystal).
Although the spectrum of incoherent radiation in crystal is
similar to one in amorphous medium, the incoherent radiation
intensity could demonstrate substantial dependence on the crystal
orientation due to the electrons' flux redistribution in the
crystal. In the present paper we apply our method of the
incoherent bremsstrahlung simulation developed earlier to
interpretation of some recent experimental results obtained at the
Mainz Microtron MAMI.
\end{abstract}

\section{Introduction}
It is well known (see, e.g. \cite{TM, AhSh, Ugg}) that high energy
electron beam incident on an oriented single crystal produces the
coherent radiation that is due to the spatial periodicity of the
lattice atoms, and the incoherent one, that is due to the thermal
spread of atoms from their positions of equilibrium in the
lattice. For the first look, the incoherent part of radiation is
similar to the last in amorphous medium (with Bethe-Heitler
spectrum), and do not depend on the crystal orientation in
relation to the particles beam.

However, in \cite{Sh2, Sh3} it was paid attention to the fact that
some features of the particle's dynamics in the crystal
(channeling effect etc.) could lead to various substantial
orientation effects in the hard range of the spectrum, where (for
$\varepsilon\sim 1$ GeV electrons) the incoherent part is
predominant. The semi-numerical approach developed in \cite{Sh2,
Sh3} was used for interpretation of early experimental data
\cite{Sanin}.

The ideas of \cite{Sh3} had been referred by the authors of recent
experiments \cite{Backe} to interpret some of their results. In
our article we present the results of simulation of the incoherent
radiation under the conditions of the experiment \cite{Backe}. A
good agreement with the experimental data confirms the
interpretation given in \cite{Backe}.

In the present report we present the results of simulation of the
incoherent bremsstrahlung under the conditions of the recent
experiment \cite{Backe2}, were the radiation from the electrons
moving in periodically bent crystal had been registered.

\section{Bremsstrahlung in dipole approximation}

Radiation of relativistic electron in matter develops in a large
spatial region along the particle's momentum. This region is known
as the coherence length (or formation length) \cite{TM, AhSh}
$l_{\mathrm{coh}} \sim 2\varepsilon\varepsilon ' /m^2c^3\omega$,
where $\varepsilon$ is the energy of the initial electron,
$\omega$ is the radiated photon frequency, $\varepsilon ' =
\varepsilon -\hbar\omega$, $m$ is the electron mass, $c$ is the
speed of light. In the large range of radiation frequencies the
coherence length could exceed the interatomic distances in
crystal:
\begin{equation}
l_{\mathrm{coh}} \gg a. \label{sst:eq1}
\end{equation}
In this case the effective constant of interaction of the electron
with the lattice atoms may be large in comparison with the unit,
so we could use the semiclassical description of the radiation
process. If, in addition to that, the electron's scattering angle
on the coherence length $\vartheta_l$ satisfies the condition
$$\vartheta_l \ll \gamma^{-1},$$
where $\gamma=\varepsilon/mc^2$ is the electron's Lorentz factor,
the dipole approximation is valid \cite{AhSh}. In this
approximation the spectral density of bremsstrahlung under
subsequent collisions on atoms could be described by the formula
\begin{equation}
\frac{dE}{d\omega} = \frac{e^2\omega}{2\pi c^2} \int_\delta^\infty
\frac{dq}{q^2}\left[ 1+
\frac{(\hbar\omega)^2}{2\varepsilon\varepsilon '} - 2
\frac{\delta}{q} \left( 1-\frac{\delta}{q} \right) \right] \left|
\sum_n \boldsymbol\vartheta_n e^{icqt_n} \right|^2,
\label{sst:eq2}
\end{equation}
where $\delta = m^2c^3\omega /2\varepsilon\varepsilon ' \sim
l_{\mathrm{coh}}^{-1}$, $\boldsymbol\vartheta_n$ is the
two-dimensional electron scattering angle under collision with the
$n$-th atom, $t_n$ is the time moment of the collision.

Consider now the radiation of the electron incident onto the
crystal under small angle $\psi$ to one of its crystallographic
axes. It is known \cite{TM, AhSh} that averaging of the value
$\left| \sum_n \boldsymbol\vartheta_n e^{icqt_n} \right|^2$ over
the thermal vibrations of atoms in the lattice leads to the split
of this value (and so the radiation intensity) into the sum of two
terms describing coherent and incoherent effects in radiation:
\begin{eqnarray}
\left< \left| \sum_n \boldsymbol\vartheta_n e^{icqt_n} \right|^2
\right> = \sum_{n,m} e^{icq(t_n-t_m)} \left< \boldsymbol\vartheta
(\boldsymbol\rho_n + \mathbf u_n) \right> \left<
\boldsymbol\vartheta
(\boldsymbol\rho_m + \mathbf u_m) \right> \\
+ \sum_n \left\{ \left< \left( \boldsymbol\vartheta
(\boldsymbol\rho_n + \mathbf u_n) \right)^2 \right> - \left(
\left< \boldsymbol\vartheta (\boldsymbol\rho_n + \mathbf u_n)
\right> \right)^2 \right\}, \label{sst:eq5}
\end{eqnarray}
where $\boldsymbol\rho_n = \boldsymbol\rho (t_n) -
\boldsymbol\rho_n^0$ is the impact parameter of the collision with
the $n$-th atom in its equilibrium position $\boldsymbol\rho_n^0$,
$\boldsymbol\rho (t)$ is the trajectory of the electron in the
plane orthogonal to the crystallographic axis (which could be
obtained by numerical integration of the equation of motion), and
$\mathbf u_n$ is the thermal shift of the $n$-th atom from the
position of equilibrium. In the range of radiation frequencies for
which
\begin{equation}
l_{\mathrm{coh}} \ll a/\psi , \label{sst:eq6}
\end{equation}
where $a$ is the distance between two parallel atomic strings the
closest to each other, the incoherent term (\ref{sst:eq5}) makes
the main contribution into the bremsstrahlung intensity
(\ref{sst:eq2}).

The radiation by the uniform beam of particles is characterized by
the radiation efficiency, that is the radiation intensity
(\ref{sst:eq2}) integrated over impact parameters of the
particles' incidence onto the crystal in the limits of one
elementary cell. So, the efficiency is the classical analog of the
quantum cross section. In the further consideration we shall
compare the radiation efficiency in the crystal to the
Bethe-Heitler efficiency of bremsstrahlung in amorphous medium.

For further computational details see \cite{Sh2, Sh3}.

\section{Origin of the orientation dependence of the incoherent bremsstrahlung}
When charged particles are incident onto the crystal under small
angle $\theta$ to one of the atomic planes densely packed with
atoms, the channeling phenomenon could takes the place (see, e.g.,
\cite{AhSh, Ugg}). Under planar channeling the electron moves in
the potential well formed by the attractive continuum potential of
the atomic plane (see figure 1, left panel). The largest incidence
angle, for which the capture into the channel is possible, is
called as the critical channeling angle $\theta_c$ \cite{AhSh,
Ugg}.

\begin{figure}[h]
\includegraphics[scale=0.5]{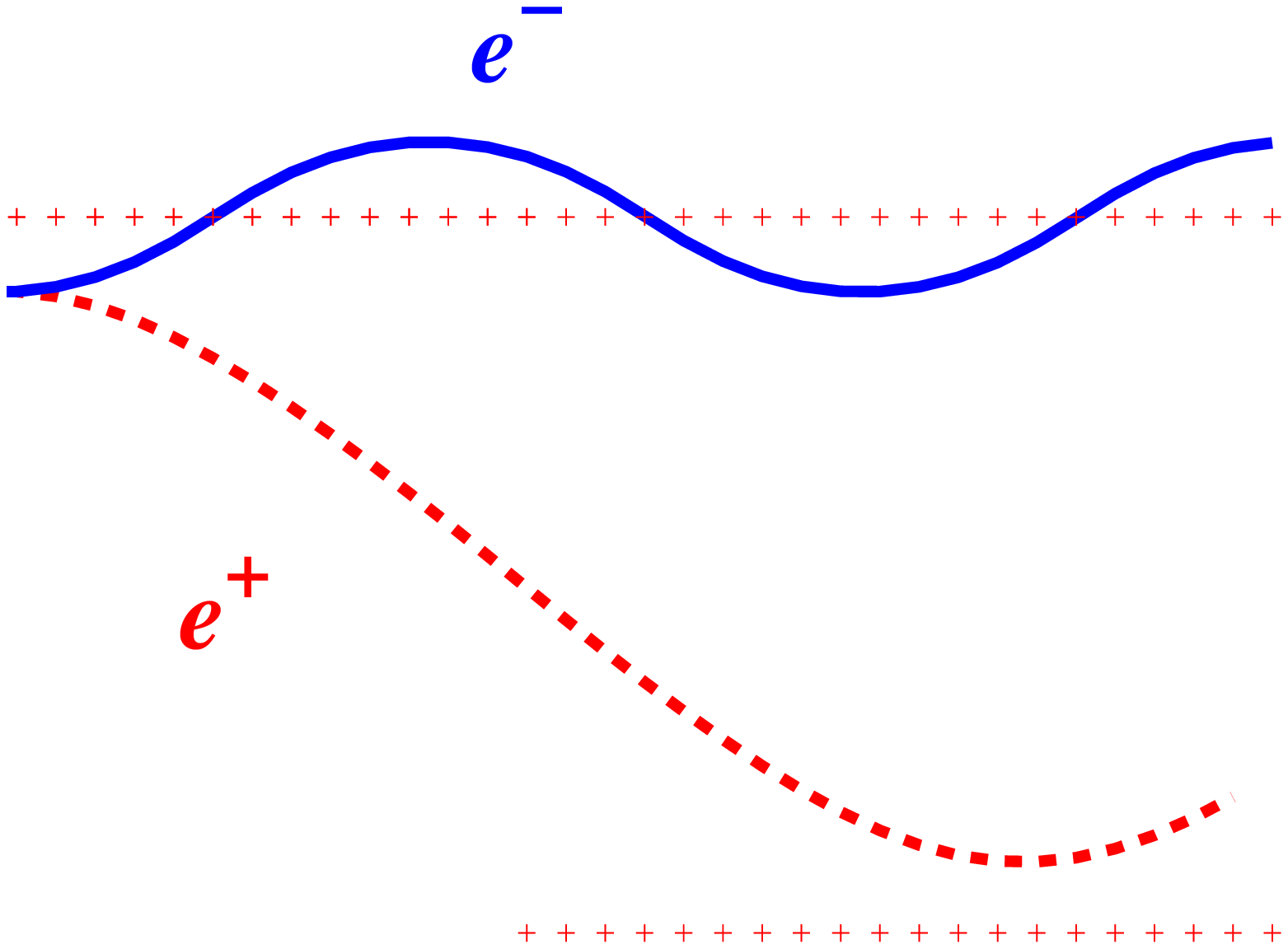} \ \
\includegraphics[scale=0.5]{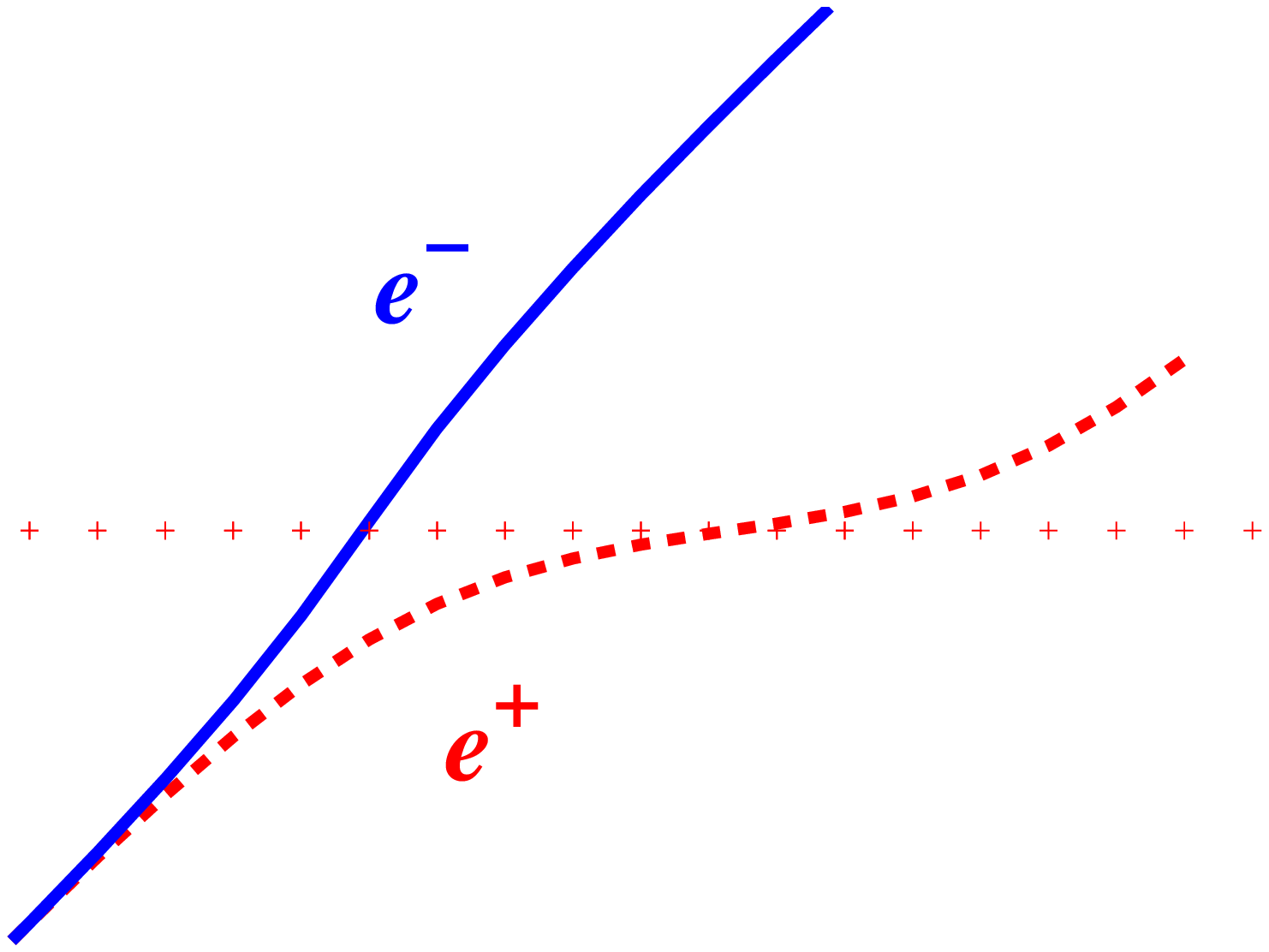}
\caption{Typical trajectories of the electrons (\full) and
positrons (\dashed) under planar channeling (left) and
above-barrier motion (right). Pluses mark the positions of atomic
strings (perpendicular to the plane of the figure) forming the
atomic planes of the crystal.}
\end{figure}

\begin{figure}[h]
\includegraphics[scale=0.55]{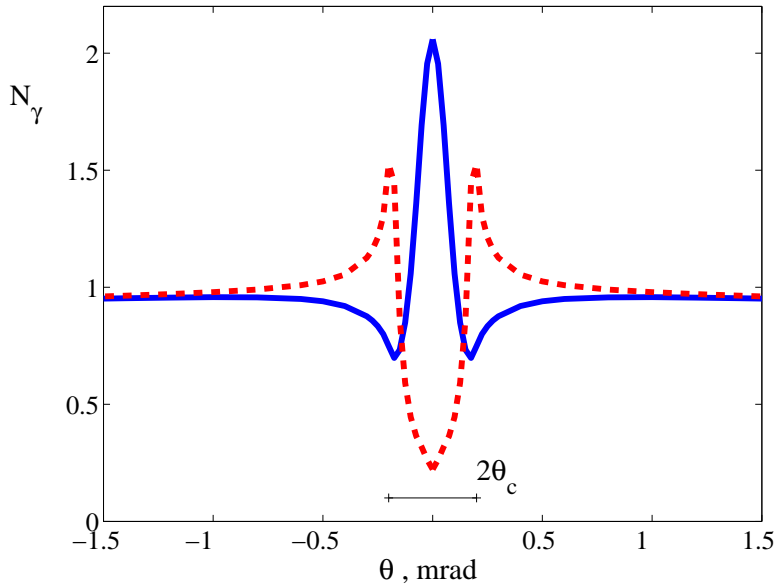}\hspace{1pc}
\begin{minipage}[b]{12pc}\caption{Incoherent bremsstrahlung efficiency (in ratio to the
Bethe-Heitler efficiency in amorphous medium) from 1 GeV electrons
(\full) and positrons (\dashed) vs incidence angle $\theta$ to
$(0\bar 11)$ plane of Si crystal, as a result of simulation.}
\end{minipage}
\end{figure}

Under $\theta\ll\theta_c$ the most part of the incident electrons
would move in the planar channeling regime. These electrons will
collide with atoms at small impact parameters more frequently then
in amorphous medium, that leads to the increase of the incoherent
bremsstrahlung efficiency (see figure 2). For $\theta\sim\theta_c$
the above-barrier motion in the continuum potential takes the
place for the most part of the particles (figure 1, right panel).
Above-barrier electrons rapidly cross the atomic plane, with
reduced number of close collisions with atoms comparing to the
case of amorphous medium. This leads to the decrease of the
incoherent bremsstrahlung efficiency (figure 2). For the positron
beam the situation is opposite. Incoherent multiple scattering on
the thermal vibrations on the lattice atoms leads to dechanneling
of the particles and, as a consequence, to smoothing of the
orientation dependence described above \cite{Sh3}.

\begin{figure}[h]
\includegraphics[width=\textwidth]{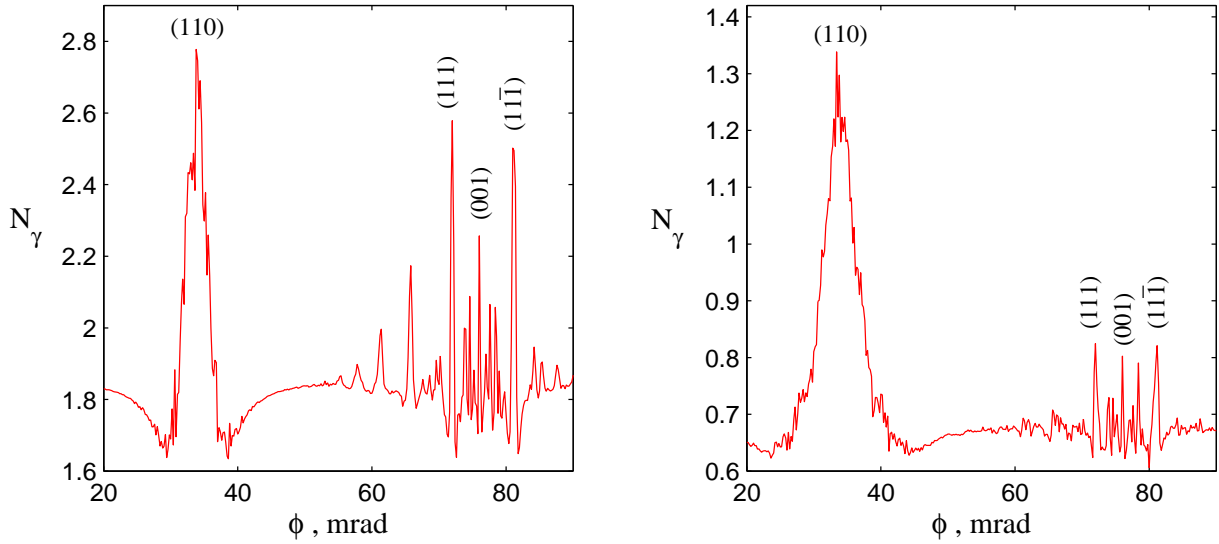}
\caption{Results of simulation for the incoherent bremsstrahlung
intensity (in ratio to Bethe-Heitler intensity in amorphous
medium) from 855 MeV electrons in flat (left plot) and
sinusoidally bent (right plot) silicon crystals under scanning of
the goniometric angle like in the experiment  \cite{Backe2}.}
\end{figure}

\begin{figure}[h]
\includegraphics[width=\textwidth]{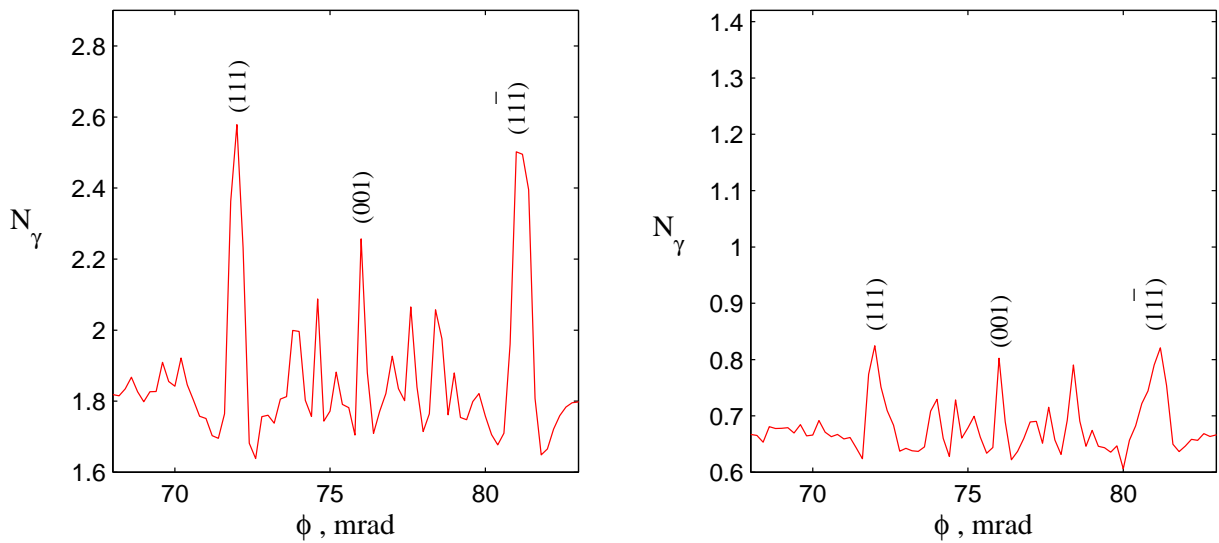}
\caption{Increased part of the previous figure.}
\end{figure}

\section{Results and discussion}
The simulation was carried out under the conditions of the recent
experiment performed at the Mainz Microtron MAMI \cite{Backe2} to
explore the radiation emission from silicon crystal with 4-period
bent (110)-planes (period of oscillations $\lambda_U=7$ $\mu m$,
amplitude $A=4.8$ \AA , electron energy $\varepsilon = 855$ MeV).
The radiation yield with the photon energy $\hbar\omega
=\varepsilon /2$, for which the incoherent radiation mechanism is
predominant, had been registered.

For the simulation we let the crystal is aligned on the goniometer
in such a way that the zero angle of incidence to (110)-plane is
achieved for the goniometer angle $\phi \approx 34$ mrad, and the
zero angle of incidence to (001)-plane is achieved for
$\phi\approx 76$ mrad, like in the experiment \cite{Backe2}.

The results of simulation (figure 3 and 4) demonstrate (at least
qualitative) agreement with the experimental data.


\end{document}